\documentstyle{mn}

\def\pmb#1{\setbox0=\hbox{#1}%
    \kern-.025em\copy0\kern-\wd0
    \kern.05em\copy0\kern-\wd0
    \kern-.025em\raise.0433em\box0 }
\def\btheta{\pmb{$\theta$}}

\newcounter{parentequation}
\def\beglet{
  \addtocounter{equation}{1}%
  \setcounter{parentequation}{\value{equation}}%
  \setcounter{equation}{0}%
  \def\theequation{\arabic{parentequation}\alph{equation}}%
  \ignorespaces
}
\def\endlet{
  \setcounter{equation}{\value{parentequation}}%
  \def\theequation{\arabic{equation}}%
}

\def\ltsima{$\; \buildrel < \over \sim \;$}
\def\gtsima{$\; \buildrel > \over \sim \;$}
\def\simlt{\lower.5ex\hbox{\ltsima}}
\def\simgt{\lower.5ex\hbox{\gtsima}}

\def\etal{{\it et al.}\rm}
\def\etals{{\it et al. }\rm}

\begin{document}

\title[Bayesian Evidence]
{Limitations of Bayesian Evidence Applied to Cosmology}

\author[G. Efstathiou]{G. Efstathiou\\
Institute of Astronomy, Madingley Road, Cambridge, CB3 OHA.}

\maketitle

\begin{abstract}
  There has been increasing interest by cosmologists in applying
  Bayesian techniques, such as Bayesian Evidence, for model selection.
  A typical example is in assessing whether observational data favour
  a cosmological constant over evolving dark energy. In this paper,
  the example of dark energy is used to illustrate limitations in the
  application of Bayesian Evidence associated with subjective
  judgements concerning the choice of model and priors.  An analysis
  of recent cosmological data shows a statistically insignificant
  preference for a cosmological constant over simple dynamical models
  of dark energy. It is argued that for nested problems, as considered
  here,  Bayesian parameter estimation can be more informative
  than computing Bayesian Evidence for poorly motivated 
  physical models.

\vskip 0.1 truein


\vskip 0.1 truein

\end{abstract}

\section{Introduction}

\vskip 0.1 truein

Bayesian techniques for estimating the posterior distributions of
cosmological parameters are now well established in astronomy (see
Lahav and Liddle, 2006, and references therein). In the last few
years, cosmologists have become increasingly interested in statistical
techniques for model selection (for an early application
see Jaffe 1996; for recent summaries see Liddle,
Mukherjee and Parkinson 2006; Trotta 2008).  This subject has a long
history and is discussed at length  in Jeffreys' classic monograph
(Jeffreys 1961). The aim of model selection is to provide a measure by
which to rank competing models. A model that is highly predictive
should clearly be favoured over a model that is not.  Model selection,
in effect, quantifies Occam's Razor by penalizing complicated models
with many parameters that need to be finely tuned to match the data. A
topical example of model selection applied to cosmology is in
assessing whether observational data favour dynamical dark energy over
a cosmological constant (for recent discussions see Szydlowski, Kurek
and Krawiec 2006; Liddle, Mukherjee, Parkinson and Wang 2006;
Sahl\'en, Liddle and Parkinson 2007; Serra, Heavens and Melchiorri
2007). This is the example that we will use in this paper.

Models can by ranked by computing the Bayesian Evidence, $E$, defined
as the probability of the data $D$ given the model $M$,
\begin{equation}
   E(M) = \int d  \btheta P(D \vert  \btheta M) 
\pi( \btheta \vert M),  \label{E1}
\end{equation}
where $\pi(\btheta \vert M)$ is the prior distribution of model
parameters \btheta $\;$ and $P(D \vert \btheta M)$ is the likelihood
of the parameters under model $M$. The ratio of the Evidences for two
models,
\begin{equation}
B_{12} = E(M_1)/E(M_2), \label{E2}
\end{equation}
also known as the Bayes factor, provides a measure with which to
discriminate between the models.  If each model is assigned equal
prior probability, the Bayes factor gives the posterior odds of the
two models. A value for $B_{12}$ of order unity indicates that there
is little to choose between the two models but a value of, say,
$B_{12} \simlt 0.01$ suggests that the data strongly favour model $2$
over model $1$\footnote{Many authors have used qualitative guidelines
  suggested by Jeffreys (1961) to interpret Bayes factors, see Section
  3.}.  The computation of Evidence can be challenging since it
requires the evaluation of an integral (\ref{E1}) over the entire
likelihood function. This can take many hours of supercomputer time if
the cost of evaluating likelihood function at a single point within a
multi-dimensional parameter space is large. Rather than compute the
Evidence, some authors have used proxies such as the Bayesian
Information Criterion, which can be computed from the maximum of the
likelihood function. The Bayesian Information Criterion (BIC) and
various information theoretic criteria for model selection are
discussed by Liddle (2004, 2007) and by Trotta (2008); these will not
be discussed in any detail in this paper which will focus on the
Evidence defined by equation (\ref{E1}).

The application of Bayesian Evidence to cosmology has not met with
uniform approval. The methodology has been attacked vigourously
recently by Linder and Miquel (2007), and defended even more
vigourously by Liddle \etals (2007). This author agrees with the
statistical analysis of Liddle \etals (2007). Nevertheless, our
conclusions are more in sympathy with those of Linder and Miquel,
namely that Bayesian Evidence is of limited use for many applications
to cosmology.

The main reason for reaching this conclusion is that the very concept
of a model is subjective in many cosmological applications. Ideally,
one would like to test a physically well motivated model, rather than
adopting a phenomenological parameterisation, but this is rarely
possible in cosmology because the underlying physics is poorly
understood. As is evident from equation (\ref{E1}) computation of
Bayesian Evidence requires assumptions concerning the prior
distributions of any parameters specifying a model. Again, in
cosmology we rarely have strong guiding principles to help us choose
priors. Bayesian Evidence is useful in situations where hypotheses are
well motivated and when there are symmetries, or other information, to
guide the choice of priors. (Specific examples are given in the
textbook by Mackay 2003). Perceived difficulties with `subjective'
choices of priors are, of course, at the heart of the long-standing
debate between `Frequentists' and `Bayesians' (see for example,
Kendall and Stuart 1979, \S21; Jaynes 2003) and this paper has nothing
new to add to this well-worn discussion. But even if one
approaches statistics from a Bayesian point of view, as this author
does, the difficulties involved in defining models and priors
must be appreciated when interpreting Bayesian Evidence.

In this paper, I will use the problem of dark energy to illustrate the
points outlined in the previous paragraph. In the next Section, I
begin with a discussion of `skater' models (Linder 2005; Sahl\'en,
Liddle and Parkinson 2007) to demonstrate the subjectivity involved in
defining a model. The skater parameterization is clearly unphysical
and should be thought of as an approximation to a more complex model
requiring more free parameters and uncertain priors. This is typical
of many parameterizations of evolving dark energy.  I will then
discuss the constraints on simple dynamical models of dark energy,
including a `thawing' field evolving in a linear potential and a
`freezing' tracker model, to illustrate problems associated with
priors. Some general comments on model selection are presented in
Section 3 and the conclusions are summarized in Section 4.

\section{Testing Dark Energy}

The discovery that the Universe is accelerating has stimulated an
enormous amount of interest in dynamical models of dark energy (see
for example, the comprehensive review by Copeland, Sami and Tsujikawa
2006).  A particularly simple class of models is based on a scalar
field $\phi$ evolving in a potential $V(\phi)$. The equation of motion
of the field is
\begin{equation}
\ddot \phi + 3H \dot \phi = - V^\prime (\phi), \label{M1}
\end{equation}
where dots denote time derivatives, $H = \dot a/a$ where $a$ is the
scale factor, and the prime denotes a derivative with respect to the field
value  $\phi$. In this paper, we focus on comparing this class
of dynamical models with the hypothesis that the dark energy is
a cosmological constant, {\it i.e.} $V = V_0 = {\rm constant}$.

\subsection{The difficulty of defining a physically well motivated model.}

In a `skating' model (Linder 2005) the potential is assumed to be
flat, $V=V_0$, but the scalar field has some kinetic energy, $\dot
\phi \ne 0$. This leads to a non-trivial equation of state that
evolves as
\begin{equation}
{dw \over d {\rm ln} a} = - 3(1-w^2). \label{M2}
\end{equation}
What prior should we choose for $\dot \phi$? The equation of motion
gives $\dot \phi \propto a^{-3}$, {\it i.e.} the field velocity decays
adiabatically as the Universe expands. So a physically well motivated
prior for $\dot \phi$ would be a delta function centred around the
value $\dot \phi = 0$, making the model indistinguishable from a
cosmological constant.

Sahl\'en, Liddle and Parkinson (2005, 2007) have derived constraints
on skater models using distant supernovae and other cosmological data.
The data do not constrain strongly the field velocity $\dot \phi_0$ at the
present day and so the limits on $\dot \phi_0$ simply reflect the
maximum value allowed by their choice of prior on the kinetic
contribution to the cosmic density at high redshift.  It can be argued
that if the likelihood function is flat over the full domain of a
parameter, then the Evidence is independent of the prior (Liddle
\etals 2007). But how does one specify the domain of a parameter?  As
discussed in the next Section, if the likelihood does vary over the
parameter range, perhaps because the data have been used to suggest
the range, then the posterior distributions of other parameters, such
as $V_0$, and the Evidence (\ref{E1}) will depend on the choice of
prior.

In fact, the problem is more serious than outlined above. We have
described the skater model here because it is easy to see that it is a
proxy for a more complicated model involving more parameters (and
priors). This is true of many simple parameterizations of dynamical
dark energy.  How could skating behaviour be realized in practice?  In
the following example \footnote{We use natural units, $c= \hbar =
  1$. The reduced Planck mass is $M_{pl} = (8 \pi G)^{-1/2} = 2.44
  \times 10^{18} {\rm GeV}$ and will be set to unity unless explicitly
  stated otherwise.}
\begin{equation}
V(\phi) =  {M^{4+\alpha} \over \phi^\alpha} + V_0, \label{M3}
\end{equation}
the first term in (\ref{M3}) is a `tracker' potential with an
attractor solution (Steinhardt, Wang and Zlatev 1999). It is therefore
easy to arrange for the field to follow the attractor solution at high
redshift and then glide on to the constant part of the potential with
some finite $\dot \phi$ at low redshift. As a specific example, assume
$V_0= 2H_0^2$, $\alpha=4$, $M^{4+\alpha} = 0.05H_0^2$, then at the
present day $\Omega_{\phi_0}=0.71$, $w_{\phi_0}=-0.97$ and
$\dot{\phi_0}/H_0 = 0.246$ (actually outside the range on $\dot
\phi_0/H_0$ permitted by the priors assumed by Sahl\'en \etals
2007). We would argue that equation (\ref{M3}) with attractor initial
conditions is a physically better motivated model than a simple flat
potential with some arbitrary choice of initial $\dot \phi$. Of
course, this is a more complicated `skater-like' model and requires
the specification of priors on {\it three} parameters $M$, $\alpha$
and $V_0$. Furthermore, as the above example shows, for reasonable
values of $\alpha$ it is difficult to get a substantial deviation from
$w_{\phi_0} = -1$. The additional parameters therefore allow models
that show deviations from the dynamics of a cosmological constant at
low redshift, but the differences are small.

\subsection{Dependence on priors}

To make contact with previous work, we first analyse the simple
phenomenological model with a constant equation of state parameter
$w_0$. Evidence computations for this model have been presented by
Liddle \etals (2006) and Serra \etals (2007), using broadly similar
data to those used here. 

We use a compilation from the following web address
{\scriptsize http://braeburn.pha.jhu.edu/$\sim$ariess/R06/Davis07\_R07\_WV07.dat},
listing redshifts, distance moduli and their errors for Type 1a
supernovae. These data were used in Davis \etals (2007) and consist of
combined data from  Wood-Vasey \etals (2007) and Riess
\etals (2007). Supernovae with redshifts less than $0.02$ were
discarded to limit systematic errors associated with local peculiar
velocities, leaving $181$ supernovae with a maximum redshift of $1.755$
(sn1997ff). Following Sahl\'en \etals (2007), in addition to the
constraints on luminosity distances from Type 1a supernovae, we add
constraints on the CMB peak shift parameter ${\cal R}$ at the redshift
of decoupling and on the baryon acoustic scale parameter $A$ at the
characteristic depth of the Sloan Digital Sky Survey (Eisenstein
\etals 2005), assuming Gaussian distributions with 
\beglet
\begin{eqnarray}
 {\cal R} (z_{dec}=1089) & =&  1.70 \pm 0.03, \label{M4a} \\
 A(z=0.35) & =& 0.474 \pm 0.017.  \label{M4b}
\end{eqnarray}
\endlet For definitions of the parameters ${\cal R}$ and $A$, and
references to the numerical values listed in (\ref{M4a}, \ref{M4b}),
we refer the reader to Wang and Mukharjee (2006) and
 Sahl\'en \etals (2007). 

Spatial curvature is assumed to be zero, thus a model is specified by
the Hubble parameter $h$ (in units of $100\;{\rm km}{\rm s}^{-1}{\rm
Mpc}^{-1}$, the cosmological matter density parameter at the present
day, $\Omega_m$, and the equation of state parameter $w_0$. To
facilitate comparison with Serra \etals (2007), we adopt the identical
flat priors for $\Omega_m$ and $h$ with ranges $0.1 \le \Omega_m \le
0.5$, $0.56 \le h \le 0.72$ (and we do not attempt to justify these
choices). For $w_0$ we will adopt a flat prior over the range
$-1 \le w_0 \le -1/3$, {\it i.e.} excluding the `phantom'
regime $w_0 < -1$, and a flat prior over the range $-2 \le w_0
\le -1/3$.

Figure 1 shows likelihood contours in the $w_0-\Omega_m$ plane
marginalised over the Hubble parameter $h$. The results are broadly
compatible with the analysis presented by Serra \etal, though the
supernova sample used here is larger and so the contours in Figure 1
are somewhat tighter than theirs. Note that the peak of the likelihood
function is close to the cosmological constant value $w_0 = -1$. There
is no evidence of a shift of the contours below the phantom divide
line seen in some earlier analyses ({\it e.g.} Riess \etal,
2004). There is evidence that the older High-z Supernovae Search
Team (HZSST) data pull the solutions to $w_0<-1$ (see Nesseris and
Perivolaroploulos 2007) indicative of (unknown) systematic errors in
the earlier data.  The HZSST data are not included in the sample used
here.

\begin{figure}

\vskip 2.9 truein

\includegraphics{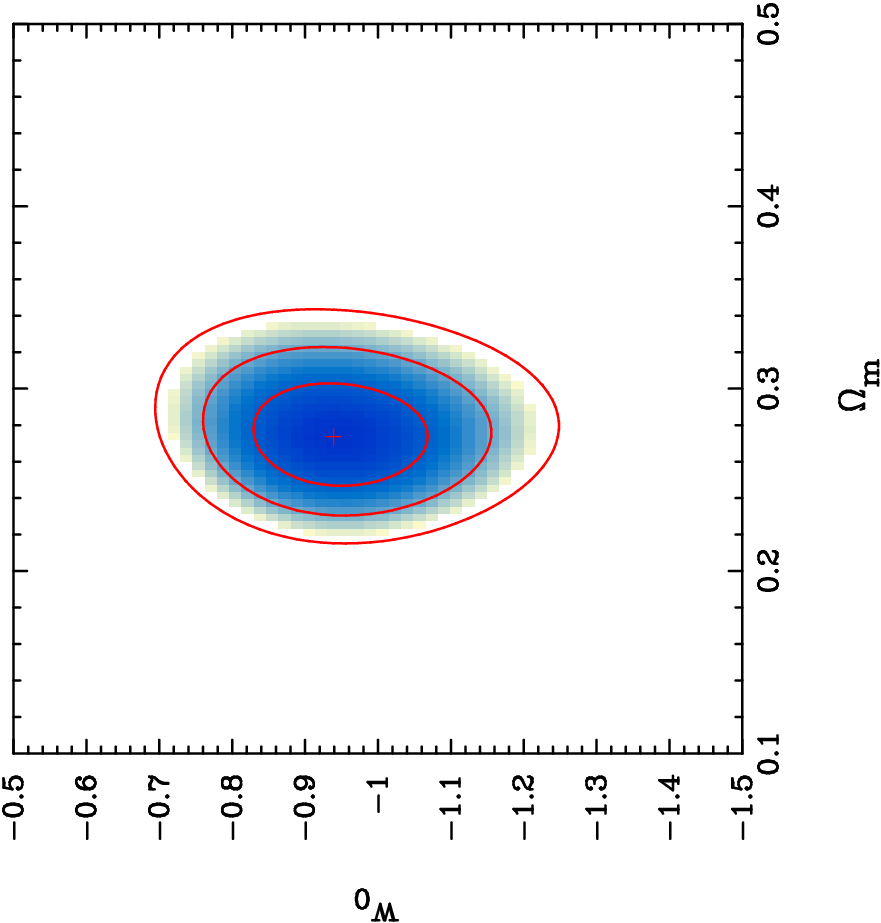}

\caption
{Constraints on $w_0$ and $\Omega_m$ from the data (primarily distant
supernovae) described in the text.  The ellipses show $1$, $2$ and
$3\sigma$ contours of the marginalized likelihood function. The
maximum of the likelihood function is shown by the cross.}
\label{figure1}

\end{figure}

The Evidence ratios for a cosmological constant and the constant $w_0$
model are listed in Table 1 for various choices of prior on $w_0$. The
results in Table 1 agree well with those of Liddle \etals (2006) who
used the distant supernovae data of Astier \etals (2006). The Evidence
ratios in Table 1 are about a factor of two larger than those computed
by Serra \etals (2006).  Much of this difference is caused because the
latter authors include the old HZSST data which pull the likelihood
function further into the phantom regime hence penalising the $\Lambda$
model.

The Evidence ratios in this Table indicate a marginal preference for
the $\Lambda$ model (see Section 3.1 for remarks on the interpretation
of Evidence), but none of the Evidence ratios are high and it is easy
to change them by factors of a few by changing the range of the prior
on $w_0$. This is demonstrated in the third line of the Table, where
the Evidence has been recomputed assuming a flat prior over the
narrower range $-1.4 \le w_0 < -0.6$. The dependence on the prior is
not particularly serious in this case, because none of the entries in
Table 1 provide strong evidence to favour or disfavour the $\Lambda$
model.

\begin{figure*}

\vskip 3.0 truein
\includegraphics{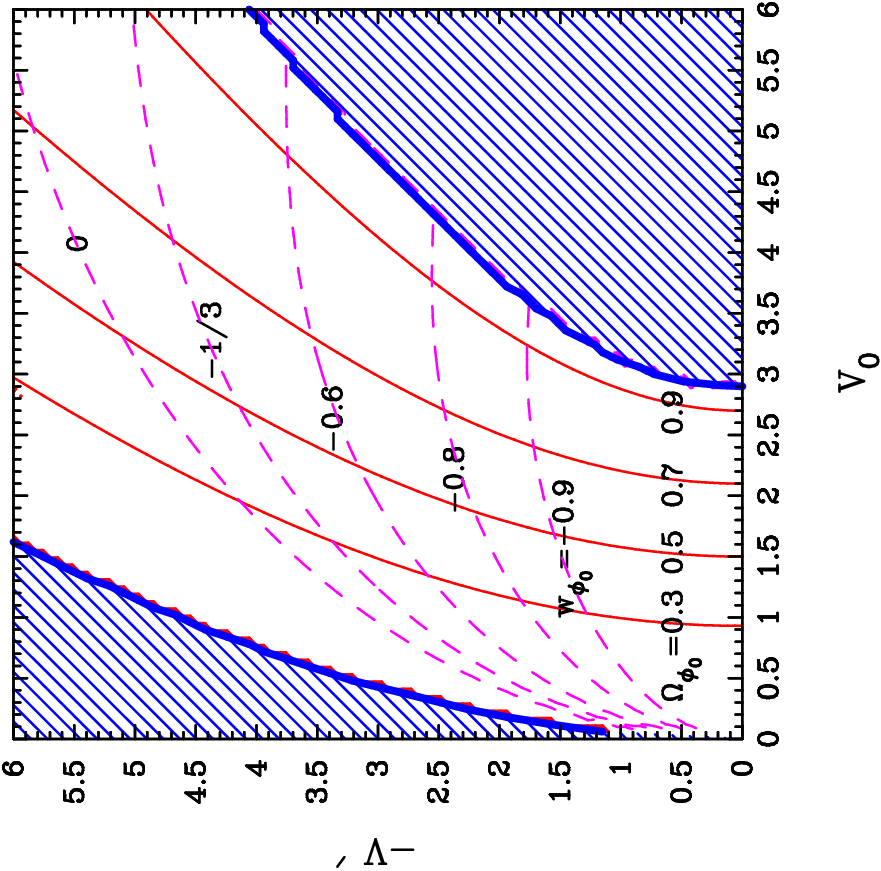}

\includegraphics{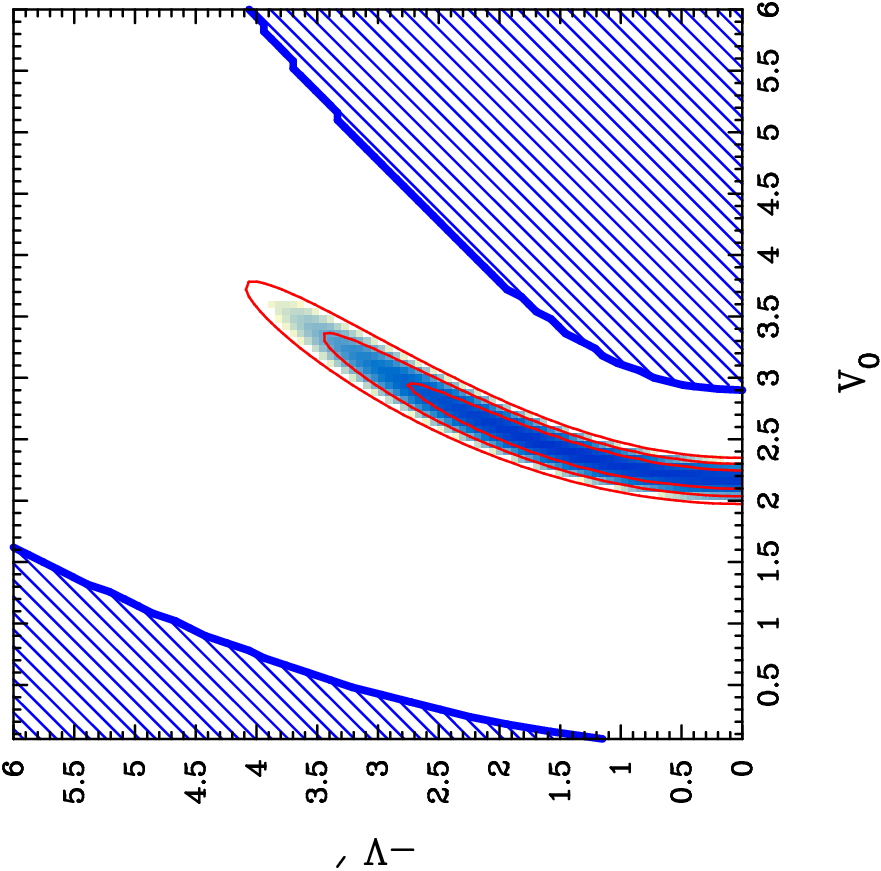}

\caption
{The Figure to the left shows regions contours of fixed values of 
$\Omega_{\phi_0}$ and $w_{\phi_0}$ 
for the linear `thawing' model discussed in the text. No solutions
are possible in the shaded regions. The Figure to the right shows the 
likelihood function determined from the data discussed in the text after
marginalising over the Hubble parameter. The contours delineate  $1$, $2$ and
$3\sigma$ confidence regions.}

\label{figure2}

\end{figure*}

\begin{table}

\centerline{\bf \ \ \  Table 1:  Evidence Ratios}

\begin{center}

\begin{tabular}{cccc} \hline \hline
\smallskip 
{\rm Model} &  {\rm Prior}  & $E_\Lambda/E_{Q}$ & ${\rm ln}(E_\Lambda/E_Q)$ \cr
\hline\hline
{\rm constant} \ $w$ & $-1 \le w_0  < -0.333$ & $3.42$ & $1.23$  \cr
{\rm constant} \ $w$ & $-2 \le w_0  < -0.333$ & $6.15$ & $1.82$  \cr
{\rm constant} \ $w$ & $-1.4 \le w_0  < -0.6$ & $2.95$ & $1.08$  \cr
{\rm linear} & $0 \le V_0 < 3$ & $1.58$ & $0.46$ \cr
{\rm linear} & $0 \le V_0 < 6$ & $2.59$ & $0.95$ \cr
{\rm inverse} \ {\rm power} & $0 \le \alpha < 6$ & $8.59$  & $2.15$ \cr
{\rm inverse} \ {\rm power} & $0 \le \alpha < 2$ & $2.78$  & $1.02$ \cr
{\rm inverse} \ {\rm power} & $0 \le \alpha < 1$ & $1.40$  & $0.33$ \cr
\hline
\end{tabular}
\end{center}
\noindent
Note: $E_\Lambda$ denoted the Evidence for a model with
a cosmological constant $\Lambda$ and $E_Q$ denotes the
Evidence for the dynamical dark energy models discussed
in the text. The first three lines list Evidence ratios
for the model with constant $w_0$. The next two lines
list Evidence ratios for the model with a linear potential 
as discussed in the text. The last three lines list Evidence
ratios for tracker models with an inverse power-law potential.
\end{table}

Following on from the discussion in Section 2.1, we would argue that a
model with a flat prior on a constant value of $w_0$ is not
particularly well motivated. We therefore seek a simple dynamical
model, derivable from a potential $V(\phi)$, with as few free
parameters as possible. One such model is based on the linear
potential\footnote{Note that the dimensionless parameters appearing in
  this equation are related to dimensional parameters as $\phi/M_{pl}
  \rightarrow \phi$, $V_0M_{pl}^{-2}H_0^{-2} \rightarrow V_0$ and
  $V^\prime M_{pl}^{-1} H_0^{-2} \rightarrow V^\prime $.},
\begin{equation}
V(\phi) = V_0 + V^\prime\phi,         \label{L1}
\end{equation}
with a negative gradient $V^\prime$.  (For discussions of the linear
potential see {\it e.g.} Dimopoulos and Thomas 2003; Kallosh \etals
2003; Avelino 2005.)  The zero point of the field value has no
significance and so can be set  to zero at some starting redshift
$z_i$. The field is locked by Hubble friction (equation {\ref{M1})
until the Hubble parameter drops sufficiently that the field begins to
roll. At this point, the dark energy will show interesting dynamical
behaviour. Eventually, the potential will become negative and the
Universe will collapse. This simple model therefore displays `thawing'
behaviour in the nomenclature of Caldwell and Linder (2005), followed
by `cosmic doomsday' in the nomenclature of Kallosh \etals (2003).

This model has a weak dependence on the starting redshift, which we
fix to be the decoupling redshift $z_i = 1089$. We `absorb' this weak
dependence into the priors on $V_0$ and $V^\prime$. A model is
therefore specified by these two parameters. The left hand panel of
Figure 2 shows contours in the $V_0 - V^\prime$ plane with constant
values of $\Omega_{\phi_0}$ (the present day dark energy density) and
$w_{\phi_0}$ (the present day dark energy equation of state
parameter).  The shaded regions delineate areas where no solution
exists.  Figure 1 shows that the data favour models with $\Omega_m
\sim 0.27$ and $w_0 \simlt -0.85$. The marginalised likelihood
function for the linear model is therefore expected to delineate a
narrow banana  centred around the $\Omega_{\phi_0} \sim 0.7$ line. 
This is indeed what is found, as illustrated in the right hand
panel of Figure 2. 

To compute the Evidence, priors need to be specified for the
parameters $V_0$ and $V^\prime$. But how do we choose these priors?
For spatially flat models with $V^\prime=0$, the value of $V_0$ must
lie within the range $0 \le V_0 < 3$. However, if we allow non-zero
values of $V^\prime$, $V_0$ can lie outside this range. One
possibility would be to choose a flat prior over the region of the
$V_0 - V^\prime$ plane corresponding to models which are accelerating
at the present day. However, this choice of prior is hardly
compelling.

In fact, the choice of prior is not particularly critical for the
interpretation of Figure 2 because current data provide relatively
poor constraints on $V^\prime$. This is illustrated by the last two
lines in Table 1 which list the Evidence ratios assuming a uniform
prior in $V^\prime$ over the full range shown in Figure 2 ($0 \le
-V^\prime < 6$) and a uniform prior in $V_0$ over the ranges $0 \le
V_0 < 3$ and $0 \le V_0 < 6$ (excluding the hatched regions). There is
no significant evidence to favour $\Lambda$ over the dynamical model,
and it is clear from inspection of Figure 2 that the higher Evidence
ratio in the last line of Table 1 is largely a consequence of the 
increased range of the prior on $V_0$.

\begin{figure}

\vskip 2.9 truein

\includegraphics{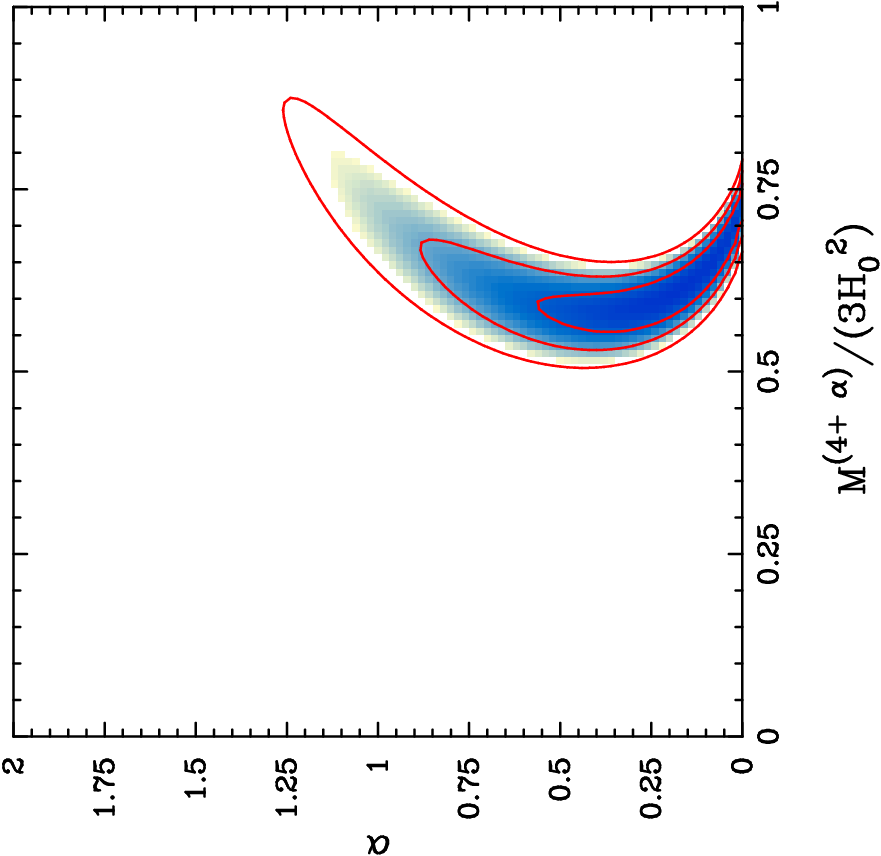}

\caption
{Constraints on the parameters of the inverse power
law potential $\alpha$ and $M$. As in Figure 1, the ellipses show $1$, $2$ and
$3\sigma$ contours of the marginalized likelihood function.}
\label{figure4}

\end{figure}

As a final example, we consider a `tracker' model with the
Ratra-Peebles (1998) potential
\begin{equation}
V(\phi) = {M^{4 + \alpha} \over \phi^\alpha} .         \label{RP1}
\end{equation}
An introductory review of this model is given by Martin (2008).  As
mentioned above, this potential has an attractor solution which drives
$w_\phi$ towards the solution
\begin{equation}
w_\phi \rightarrow {\alpha w_B - 2 \over \alpha +2}, \label{RP2}
\end{equation}
while the scalar field is subdominant (where $w_B$ is the equation of
state parameter of the background matter). This model is an 
example of a `freezing' model in the terminology of Caldwell and Linder (2005).
We set $\dot \phi = 0$ at
high redshift ($z = 10000$) and choose the initial value of $\phi$ so
that the field locks on to the attractor solution without
overshoot. The low redshift behaviour is therefore fixed by the
attractor solution so the model is characterised by the two parameters
$M$ and $\alpha$ defining the potential (\ref {RP1}).

The marginalised likelihood for this model is shown in Figure 3,
(using the same observational data as for Figures 1 and 2). Evidently,
the data constrain the power-law index to be $\alpha \simlt
1$. However, from the theoretical point of view, there are no
compelling constraints on the spectral index; values as high as
$\alpha = 6$ or more have been discussed in the literature
(Steinhardt, Wang and Zlatev 1999, Martin 2008) and occasionally
promoted as a possible solution to the `hierarchy' problem associated
with the mass scale $M$.  The Evidence ratios for various assumed
prior ranges in $\alpha$ are listed in the last three lines of Table
1. As expected, these scale almost perfectly with the width of the
prior range assumed for $\alpha$.  Nevertheless, none of the Evidence
ratios in Table 3 are high and so one cannot reject the potential
(\ref{RP1}) drawn from a broad uniform prior on $\alpha$. However, it
is obvious from the likelihood function shown in Figure 3 that we can
rule out the potential (\ref{RP1}) for {\it specific} choices of 
$\alpha \simgt 1.5$.

\section{Comments on the Application of Model Selection}

\subsection{The Jeffreys Scale}

The previous Section shows that current data, unfortunately, provide
relatively weak constraints on simple models of dynamical dark energy.
The highest Evidence ratios are about $\Delta {\rm ln}E \sim 2$ for certain
choices of prior. Plausible variations on the prior of a single
parameter can easily change the Evidence ratio by $\Delta {\rm ln} E \sim 1$
or more. Many papers on cosmological model selection have adopted the
interpretive scale suggested in Appendix B by Jeffreys, which in the
view of this author is not sufficiently conservative. Trotta (2008)
presents a revised interpretative scale in his review, which accords
with ones intuitive assessment of the relative posterior odds,
$B_{12}$, of two models. The two scales are compared in Table 2.

\begin{table}

\centerline{\bf \ \ \  Table 2:  Interpretive scales}

\begin{center}

\begin{tabular}{cc||cc} \hline
\multicolumn{2}{|c|}{\rm Jeffreys grades} & \multicolumn{2} 
{|c|}{\rm Trotta (2008)} \\
\smallskip 
${\rm ln}B_{12}$ &  {\rm strength of Evidence}  & ${\rm ln}B_{12}$ &  {\rm strength of Evidence}  \\ \hline
 $<1.15$ & {\rm not worth a mention} & $<1.0$ & {\rm inconclusive}  \cr
$1.15-2.3$ & {\rm substantial} & $1.0-2.5$ & {\rm weak}  \cr
$2.3 -4.6$ & {\rm strong to very strong}  & $2.5-5.0$ & {\rm moderate to strong} \cr
$> 4.6$ & {\rm decisive}    & $> 5$ & {\rm strong} \cr
\hline
\end{tabular}
\end{center}

\end{table}

For the dark energy examples summarized in the previous Section, if
the observational data improve to give $\Delta {\rm ln} (E_\Lambda/E_Q)
\simgt 5$, then it would be reasonable to conclude that there is
evidence favouring a cosmological constant. The data are then
providing strong enough constraints to overwhelm the changes in the
prior volumes illustrated in Table 1 (see Section 3.3 below). But
Evidence ratios of $\Delta {\rm ln} E \sim 2$ are clearly too small to
achieve this. Many of the problems outlined in the previous Section
can be overcome by adopting a conservatively high threshold before
claiming strong evidence against particular classes of model. The
threshold may need to be set higher than $\Delta {\rm ln} E = 5$ if one (or
both) of the models involves several additional parameters with
uncertain priors.

\subsection{Stating and Varying Priors}

It is essential that authors computing Bayesian Evidence state their
priors carefully since these are an integral part of the definition of
a model. If there are no compelling reasons to guide the choice of
priors, then one should demonstrate that the data overwhelm plausible
variations in priors before drawing any strong conclusions on
particular classes of model.  This has not been common practice in the
literature. For example, Table 4 of Trotta (2008) summarizes Evidence
calculations for various cosmological model comparisons. Of the ten
entries testing dynamical dark energy agains $\Lambda$, only three
explore variations in priors. One analysis quoted in this Table
(Bassett, Corasaniti and Kunz 2004) computes Evidence for simple
parameterizations of $w$ (such as $w = w_0 + w_1z$) without stating
the prior ranges on the parameters. These authors find $\Delta
E_\Lambda/E_Q \simgt 5-6$, suggesting strong evidence favouring
$\Lambda$. Such high Evidence ratios are 
 clearly at variance with the conclusions of Section 2.

\subsection{Model Selection compared with Parameter Estimation}

In many cosmological applications of model selection, we are dealing
with highly nested problems. In each of the examples discussed in
Section 2, the model for dynamical dark energy  tends to the
$\Lambda$ model as one additional the parameter tends to zero ($w_0+1
\rightarrow 0$, $V^\prime \rightarrow 0$, $\alpha \rightarrow 0$). In
each of these cases, the primary question of interest is whether there
is any empirical evidence that an additional parameter, $\lambda$
differs from zero.  For such highly nested problems\footnote{ To
  simplify the following discussion, we will assume uniform priors on
  all parameters.}  the Bayes factor for model $M_1$ ($\lambda=0$) and
model $M_2$ ($\lambda$ drawn from a prior distribution $\pi(\lambda)$)
is simply
\begin{equation}
 B_{12} =  {P(D \vert \lambda=0 M_2) \over \int P(D \vert \lambda M_2) \pi(\lambda \vert M_2) d\lambda }, \label{SD1}
\end{equation}
(this is related to the Savage-Dickey density ratio, see Trotta 2007)
where $P(D \vert \lambda M_2) \equiv {\cal L}(\lambda)$ is the
likelihood function marginalised over all of the common
parameters). For a uniform prior in $\lambda$, ${\cal L}(\lambda)$ is
just the marginalised posterior distribution of 
$\lambda$ on $M_2$, and from (\ref{SD1}) we can interpret the
likelihood ratio ${\cal L}(0)/{\cal L}(\lambda)$ as the Bayes factor
for two models with delta function priors centred at $\lambda = 0$ and
$\lambda$. If we choose $\lambda = \lambda_*$, where $\lambda_*$
corresponds to the peak of the likelihood function, then the Bayes
factor is minimised. (For further discussion of the relationship
between likelihood ratios and Bayes factors see Gordon and Trotta,
2007.) Now suppose that the likelihood function is approximated
by a Gaussian
\begin{equation}
 {\cal L}(\lambda)  =  {\cal L}_0 \; {\rm exp} \left ( -{ (\lambda - \lambda_*)^2 \over 2 \sigma^2} \right ),  \label{SD2}
\end{equation}
and that we are interested in whether the parameter $\lambda$ differs from zero.
The likelihood ratio ${\cal L}(0)/{\cal L}(\lambda_*)$ is evidently
\begin{equation}
 {{\cal L}(0) \over {\cal L}(\lambda)} =  {\rm exp} \left ( -{ \lambda_*^2 \over 2 \sigma^2} \right ),  \label{SD3}
\end{equation}
Now assume that under model $2$, the parameter $\lambda$ is drawn from
a uniform distribution in the range $0 \le \lambda < \lambda{\rm
  max}$, and assume further that $\lambda_{\rm max} \gg \lambda_* \gg
\sigma$. The data is then `informative' ($\lambda_{\rm max} \gg
\sigma$) and suggestive that $ \lambda$ deviates from zero ($\lambda_*
\gg \sigma$). In this case,
\begin{equation}
 {E(0) \over E(\lambda)} \approx  {\rm exp} 
\left ( -{ \lambda_*^2 \over 2 \sigma^2} \right )
 {\lambda_{\rm max} \over \sqrt {2 \pi} \sigma} .  \label{SD4}
\end{equation}
If the exponential term dominates in (\ref{SD4}),  the Evidence ratio
is exponentially suppressed and the data swamp the dependence on the prior.
(In fact, in testing whether a parameter differs from zero, Jeffreys (1961,
\S5.2) suggests using the prior
\begin{equation}
 \pi(\lambda) \propto {1 \over (1 + \lambda^2/\sigma^2)}, \label{SD4a}
\end{equation}
since `there is nothing in the problem except $\sigma$ to give a scale
for $\lambda$'. In other words, the choice of prior is driven by the data
and in this case one is reliant on the exponential factor (\ref{SD4}) to
reject the null hypothesis.)
Alternatively we might find $\lambda_* \simlt \sigma$ suggesting that the
parameter $\lambda$ is consistent with zero, with $\lambda_{\rm max} \gg \sigma$,
in which case 
\begin{equation}
{E(0) \over E(\lambda)} \approx  {2 \over \sqrt 2 \pi} {\lambda_{\rm max} \over \sigma},  \label{SD5}
\end{equation}
which merely tells that under model $2$ we need fine tuning of order $\sigma/\lambda_{\rm max}$ to explain the data.

\begin{table}

\centerline{\bf \ \ \  Table 3:  Likelihood ratio compared to Evidence ratio}

\begin{center}

\begin{tabular}{cccc} \hline
\multicolumn{2}{|c|} {$\;$ } & \multicolumn{2} 
{|c|}{$ {\rm ln} (E_\Lambda/E_Q)$ {\rm uniform prior}} \\
\smallskip 
 $\sigma$ &  ${\rm ln} {\cal L}(-1)/{\cal L}(w_*)$ &   $-1 \le w < -1/3$  &  $-1 \le w < -0.90$   \\ \hline
$0.1$ & $-0.03$ & $1.22$ & $-0.084$  \cr
$0.05$ & $-0.50$ & $1.77$ & $-0.021$  \cr
$0.02$ & $-3.12$ & $-0.53$ & $-2.42$  \cr
$0.015$ & $-5.55$ & $-2.67$ & $-4.58$  \cr
$0.01$ & $-12.50$ & $-9.22$ & $-11.11$  \cr

\hline
\end{tabular}
\end{center}

\end{table}

As a specific example\footnote{We could equally as well have used the
  example of the spatial curvature $\Omega_k$ or deviation of the
  scalar spectral index $n_s$ from unity,  as discussed by Liddle
  \etals (2007).}, imagine that future dark energy surveys find $w_* =
-0.95 \pm \sigma$ ($w_0 \equiv (1-\lambda)$ in the notation used
above) and we test model $1$ ($w_0 = -1$, the `null hypothesis')
against model $2$ (uniform prior in $w_0$). Table 3 lists the
likelihood ratios and Bayes factors for two choices of prior. One can
see that for $\sigma = 0.01$ the data swamp the dependence on the
prior and all three entries in Table 3 strongly disfavour $w_0 =
-1$. The likelihood ratio is as informative as the Evidence ratios in
the exponentially dominated regime since it matters little whether the
posterior odds of two models are $\sim 10^{-6}$ or $\sim 10^{-10}$ --
the odds are negligible in either case . The case of $\sigma = 0.015$
is more interesting and is an example of `Lindley's paradox' (Lindley
1957). The likelihood function indicates a $3.33\sigma$ discrepancy
with the null hypothesis, yet the Evidence ratio in the third column
of Table 3 suggests weak evidence against $w_0=-1$. This is simply
because the Evidence ratio compares one model that is disfavoured by
the data ($w_0=-1$) against another model that is disfavoured by the
data (uniform distribution of $\lambda$ over the range $-1 \le w <
-1/3$, requiring fine tuning at the few percent level). Lindley's
paradox should not obscure the fact {\it that the likelihood peaks
  away from} $w_0=-1$: the likelihood function suggests that $w_0$
differs from $-1$. and is therefore informative. If, following future
experiments, the contours shown in Figures 1-3 tighten so that zero
values for $1+w_0$, $V^\prime$, or $\alpha$, are exponentially
suppressed, then we will have very strong evidence in favour of
dynamical dark energy, irrespective of priors.  If the Evidence ratios
for reasonable choices of priors are still in the `ambiguous' range
${\rm ln}{E_\Lambda/E_Q} \sim 2.5-5$, a modest improvement of the the data
could potentially render the issue decisive.

Finally, let us consider the case relevant to equation (\ref{SD5}),
namely that the likelihood function peaks at $\lambda \sim 0$ to within
$\sim \sigma$. The null hypothesis is then favoured if the prior range
$\lambda_{\rm max} \gg \sigma$.  But the Evidence ratio can be reduced
to unity by adjusting the prior range to be of order $\sigma$ ({\it
  c.f} equation \ref{SD4a}). An `Occam's Razor' penalty for a model
with additional parameters can be realised only if: (a) one has good
arguments for choosing the prior ranges of the additional parameters
and (b) the likelihood function is compact with respect to these prior
ranges.

\subsection{When is Bayesian Evidence Particularly Useful?}

In the previous sub-section we have argued that the marginalised
likelihood function is informative and can provides a good indicator
of whether the null hypothesis ($\Lambda$) is disfavoured compared to
dynamical energy models. Let us suppose that $\Lambda$ is indeed
disfavoured by future data. In this case, the likelihood contours in
Figures 1-3 would break up into `islands' peaked away from
$\Lambda$. How do we assess between these three parameterizations?
Bayesian Evidence is likely to be indispensible for this type of
`non-nested' model comparison. Again, the usual caveats should apply:

\smallskip

\noindent
(a) we should aim to compare physically well-motivated models
(better motivated than the skater model of Section 2.1);

\smallskip

\noindent
(b) compare models with as few free parameters as possible
({\it cf} the models of Section 2.2) to limit the sensitivity
of the Evidence to prior volumes;

\smallskip

\noindent
(c) explore variations in the priors.

\section{Conclusions}

Bayesian inference has been applied widely for parameter estimation
from cosmological data and is relatively uncontroversial. The Bayesian
framework can easily be extended to model selection, but this has
proved to be more controversial.

There is nothing wrong with the mathematical framework underlying
Bayesian model selection. It is the perceived usefulness of the
framework, given difficulties in specifying models that is at the
source of the controversy.  It is important to recognise that Bayesian
model selection differs in a fundamental way from Bayesian parameter
estimation. In parameter estimation, the posterior distribution on a
parameter is useful because it usually become narrower as the quality
of constraining data improves.  The sensitivity of the posterior
distribution to the prior therefore often diminishes dramatically with
better data. This is why Bayesian parameter estimation has proved
relatively uncontroversial. (Surprisingly so, since in cases such as
estimating the CMB quadrupole there is an irreducible sensitivity to
the choice of prior, see Efstathiou 2003).

For Bayesian model selection we need to apply `physical intuition' to
select suitable models and the priors on model parameters. Once these
are chosen, the data determine the numerical value of the Evidence
({\it i.e.} the probability of the data given the model) via equation
(\ref{E1}). There is no `updating' involved since the data return a
single value of the Evidence given the model. In Section 2, we
discussed some of the difficulties associated with defining physically
well motivated models and parameter ranges.  Now one can argue,
correctly, that the range of a parameter is part of the definition of
a model. However, in cosmology, it is often difficult to provide
compelling arguments in favour of a particular parameter range.  This
is certainly the case for the dark energy tests described in this
paper. If we use the data to suggest the parameter ranges, for
example, by examining the likelihood function, then a computation of
(\ref{E1}) will be of limited value since the probability of the data
given the model will be high by construction.

Many applications of Bayesian Evidence to cosmology involve highly
nested problems in which the primary question of interest is whether a
key parameter, $\lambda$, differs from zero (the `null'
hypothesis). For such problems, we have argued that the marginalized
likelihood function ${\cal L}(\lambda)$ is more informative than
Evidence computed for specific, and often poorly motivated, choices of
priors.  If the likelihood function is exponentially suppressed at
$\lambda=0$, then we can conclude that there is strong evidence
against the null hypothesis for any reasonable choice of priors.
Bayesian Evidence is of most use is in comparing non-nested models.
If we are in the happy situation of having high quality data that
rule out a cosmological constant, then Bayesian Evidence can be
used to select between various dynamical models. But the Evidences
will only be of interest if the models and priors are physically
well motivated.

Finally, we re-iterate that the Evidence calculations presented in
Table 1 show no significant evidence in favour of a cosmological
constant compared to the dynamical models of dark energy tested here.
This conclusion agrees with similar Evidence analyses of Serra \etals
(2007) and Liddle \etals (2007), using different models and somewhat
different data,  but disagrees with the Evidence analysis of Bassett
\etals (2004).  Several recent papers have used the Bayesian
Information Criterion (BIC) to claim that a cosmological constant is
favoured over dynamical dark energy ( Bassett \etals 2004, Davis
\etals 2007; Sahl\'en \etals 2007; Kurek and Szydlowski
2007). However, BIC unfairly penalises models with many parameters if
these parameters are poorly constrained by the data (Liddle 2004;
Liddle 2007). If this paper is a `health warning' concerning the use
of Bayesian Evidence in cosmology, it should be considered a `death
certificate' on the use of approximations such as BIC if the strict
criteria for their applicability are not met (see Liddle 2007 for
further details).  As the models of Section 2 show, current data
unfortunately provide relatively weak constraints on simple dynamical
models of dark energy.

\vskip 0.1 truein

\noindent {\bf Acknowledgments:} I thank Antony Lewis for helpful
comments. The preprint version of this paper generated some lively
correspondence.  I thank the referee, Chris Gordon, Mike Hobson,
Andrew Jaffe, Eric Linder, Roberto Trotta, and particularly Andrew Liddle, for
their critical remarks which I hope have led to an improvement in the
revised version. This does not, of course, imply that they share the
perspective outlined here.

\end{document}